\documentclass[aps, pra, showpacs,
superscriptaddress, unsortedaddress, twocolumn, preprintnumbers
]{revtex4}

\usepackage{graphicx}
\usepackage{calc}
\usepackage{amsmath}
\usepackage{amssymb}

\newcommand{\Tr}{\mathop{\mathrm{Tr}}\nolimits}
\newcommand{\ket}[1]{\left|\mathrm{#1}\right\rangle}
\newcommand{\bra}[1]{\left\langle\mathrm{#1}\right|}

\newcommand{\Laguerre}{\mathop{\mathrm{L}}\nolimits}

\newcommand{\HyperGeom}{\mathop{\mathrm{F}}\nolimits}

\begin{document}
\preprint{PHYSICAL REVIEW A {\bf 78}, 055803 (2008); {\bf 79},
019902(E) (2009).}
\author{A.A. Semenov}
\email[E-mail address: ]{sem@iop.kiev.ua}  \affiliation{Institute of
Physics, National Academy of Sciences of Ukraine, Prospect Nauky 46,
UA-03028 Kiev, Ukraine}\affiliation{Institute of Physics and
Technology, National Technical University of Ukraine ``KPI'',
Prospect Peremohy 37, UA-03056 Kiev, Ukraine}\affiliation{
Bogolyubov Institute for Theoretical Physics, National Academy of
Sciences of Ukraine,\\ Vul. Metrologichna 14-b, UA-03680 Kiev,
Ukraine}

\author{A.V. Turchin}
\affiliation{Institute of Physics, National Academy of Sciences of
Ukraine, Prospect Nauky 46, UA-03028 Kiev,
Ukraine}\affiliation{Institute of Physics and Technology, National
Technical University of Ukraine ``KPI'', Prospect Peremohy 37,
UA-03056 Kiev, Ukraine}

\author{H.V. Gomonay}
\affiliation{Institute of Physics and Technology, National Technical
University of Ukraine ``KPI'', Prospect Peremohy 37, UA-03056 Kiev,
Ukraine}\affiliation{ Bogolyubov Institute for Theoretical Physics,
National Academy of Sciences of Ukraine,\\ Vul. Metrologichna 14-b,
UA-03680 Kiev, Ukraine}

\title{Detection of quantum light in the presence of noise}

\date{\today}

\begin{abstract}

Detection of quantum light in the presence of dark counts and
background radiation noise is considered. The corresponding positive
operator-valued measure is obtained and photocounts statistics of
quantum light in the presence of noise is studied.

\end{abstract}

\maketitle

Photodetectors play a crucial role in all experimental
investigations dealing with quantum optics, fundamentals of quantum
physics, and quantum-information processing. These devices are used
for measuring the photon number of radiation fields. The theory of
photodetection has been developed for both classical \cite{Mandel1,
MandelBook} and quantum light fields \cite{MandelBook, Glauber1,
Glauber2, Kelley, VogelBook}.

Different kinds of losses are a serious problem in the photoelectric
detection of quantum light. Presently available technologies enable
us to get the detection efficiency near 0.9 and even more
\cite{Waks}. At the same time, attempts to improve it increase the
dark counts rate \cite{Takeuchi}. Besides, in many applications the
background radiation noise contributes to the total statistics of
photocounts similarly to dark counts~\cite{Karp, Pratt}.

In some models (see e.g. \cite{Rohde}) noise counts, originated from
dark counts and background radiation, are described by coupling the
radiation field to a single mode of the thermal bath. However, such
a simple model has, at least, two serious drawbacks. First, it does
not predict Poissonian statistics usually \cite{Pratt} peculiar to
noise counts. Second, it does not consider the presence of other
modes of the thermal bath. Although some of these modes are not
coupled to the mode of the radiation field, they contribute to the
total statistics of photocounts. Therefore, a more appropriate model
should include a multimode noise.

For the classical fields a similar problem was considered in
Refs.~\cite{Karp, Pratt}. At the same time, quantum radiation
demonstrates many nonclassical properties such as sub-Poissonian
statistics \cite{Short}, quadrature squeezing \cite{Slusher}, etc.
These properties can be described only in the framework of the
consistent quantum formalism, which is the subject of this paper.

According to the quantum measurement theory, see, e.g.,
\cite{Busch}, the process of photodetection is characterized by the
positive operator-valued measure (POVM), $\hat{\Pi}_n$. For the
quantum radiation field characterized by the density operator
$\hat{\varrho}$, the probability distribution $P_n$ to get $n$
photocounts is written as
\begin{equation}
P_n=\Tr\left(\hat{\Pi}_n \hat{\varrho}\right).\label{POVMDefinition}
\end{equation}
A well-known result of the photodetection theory \cite{Mandel1,
MandelBook, Glauber1, Glauber2, Kelley, VogelBook} is the expression
for the POVM of a single-mode radiation field and for the detectors
with losses,
\begin{equation}
\hat{\Pi}_n=: \frac{\left(\eta\hat{a}^\dag\hat{a}\right)^{n}}{n!}
e^{-\eta\hat{a}^\dag\hat{a}}:,\label{POVMLosses}
\end{equation}
where $\eta$ is the efficiency of detection, $\hat{a}^\dag$ and
$\hat{a}$ are creation and annihilation operators of the field mode,
correspondingly, and $::$ means normal ordering. Expression
(\ref{POVMLosses}) plays a crucial role in various investigations
dealing with quantum optical measurements; see, e.g.,
\cite{Wallentowitz, Vogel, VogelPaper}. We generalize this
expression for the case of detection in the presence of noise
counts.

Let us consider the normal ordered (Husimi-Kano) symbol
\cite{HusimiKano} of the POVM,
\begin{equation}\label{Husimi}
\Pi_n\!\left(\alpha\right)=\bra{\alpha}\hat{\Pi}_n\ket{\alpha},
\end{equation}
where $\ket{\alpha}$ is a coherent state. It is obvious that
$\Pi_n\!\left(\alpha\right)$ is a probability to get $n$ photocounts
when the detector is irradiated by a coherent light with an
amplitude $\alpha$. The operator form of the POVM can be obtained
from the explicit form of expression (\ref{Husimi}) by replacing
$\alpha$ with $\hat{a}$ and $\alpha^\ast$ with $\hat{a}^\dag$ under
the sign of normal ordering.

As was mentioned above, noise counts for realistic photodetection
should be described with a multimode heat bath. Finite detection
time enables one to restrict consideration to a discrete set of
modes for the radiation field and, consequently, for the thermal
bath. The number of thermal modes $\mu$ can be approximately
evaluated as follows:
\begin{equation}
\mu\approx\Delta \omega T,\label{Mu}
\end{equation}
once the bandwidth of heat bath  $\Delta \omega$ and detection time
$T$ is known.

\begin{figure}[ht!]
\includegraphics[clip=,width=0.95\linewidth]{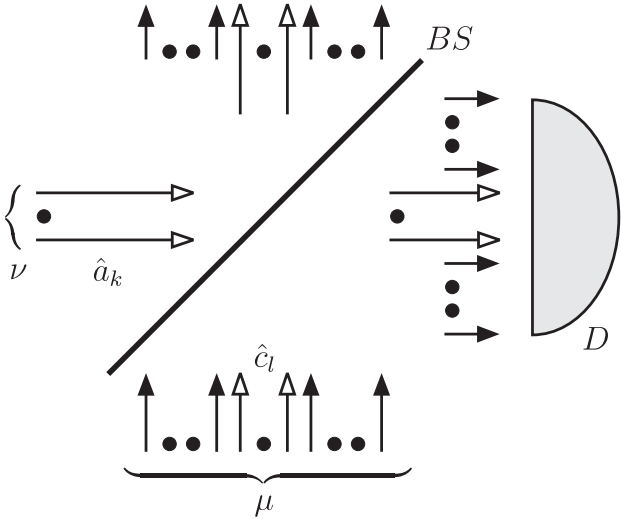}
\caption{\label{Fig_Model} The beam-splitter model for the detection
in the presence of noise. $D$ is an ideal detector, $BS$ is a beam
splitter with transmission coefficient  $\sqrt{\eta}$; the operators
$\hat{a}_{k}$, $k=1..\nu$ and $\hat{c}_l$, $l=1..\mu$ describe the
modes of the incident radiation field and thermal bath,
correspondingly.}
\end{figure}

For a sufficiently small detection time, the model with single-mode
thermal noise \cite{Rohde} seems meaningful. However, an account of
an additional thermal mode may significantly change the statistics
of photocounts; see \cite{MandelBook, Mandel2, Mandel3}. We simulate
the detection in the presence of noise counts by $\mu$ modes of the
thermal bath coupled to $\nu$ modes of the radiation field; see the
beam-splitter replacement scheme in Fig.~\ref{Fig_Model}. The
corresponding normal-ordered symbol of the POVM is equal to the
probability distribution of photocounts for mixture of the coherent
and thermal fields. The analytical expression for the probability
distribution of photocounts for the superposition of the coherent
and thermal light can be obtained under the assumption that all the
modes of the thermal bath have an equal mean number of photons,
$\bar N_\mathrm{nc}/\mu$, \cite{Saleh}, \setlength\arraycolsep{0pt}
\begin{eqnarray}
\Pi_n&&\left(\left\{\alpha_k\right\};\bar
N_\mathrm{nc},\mu\right)=\frac{\left(\frac{\bar
N_\mathrm{nc}}{\mu}\right)^n}{\left(1+\frac{\bar
N_\mathrm{nc}}{\mu}\right)^{n+\mu}}
\label{POVMNuModesSymbol}\\
&&\times\exp\left(-\frac{\eta}{1+\frac{\bar
N_\mathrm{nc}}{\mu}}\sum\limits_{k=1}^{\nu}
\left|\alpha_k\right|^2\right)\nonumber
\\ &&\times\Laguerre_n^{\mu-1}\!\left(-\frac{\eta}
{\frac{\bar N_\mathrm{nc}}{\mu}\left(1+\frac{\bar
N_\mathrm{nc}}{\mu}\right)} \sum\limits_{k=1}^{\nu}
\left|\alpha_k\right|^2\right),\nonumber
\end{eqnarray}
where $\alpha_k$, for $k=1\ldots\nu$, is the complex coherent
amplitude for the $k\textrm{th}$ mode of the radiation field, $\bar
N_\mathrm{nc}$ is the overall number of thermal photons (noise
counts) in the output of the beam splitter, and
$\Laguerre_n^{m}\!\left(x\right)$ is the generalized Laguerre
polynomial.

For the sake of simplicity, we consider a single-mode radiation
field, i.e., $\nu=1$. All the results can be generalized simply for
the case of an arbitrary $\nu$. In this case the POVM, cf.
Eq.~(\ref{POVMNuModesSymbol}), is rewritten as
\begin{eqnarray}
&&\Pi_n\left(\alpha;\bar
N_\mathrm{nc},\mu\right)=\frac{\left(\frac{\bar{N}_\mathrm{nc}}{\mu}\right)^n}
{\left(1+\frac{\bar{N}_\mathrm{nc}}{\mu}\right)^{n+\mu}}
\label{POVMMultimodeSymbol}\\
&&\times\exp\left(-\frac{\eta}{1+\frac{\bar{N}_\mathrm{nc}}{\mu}}\left|\alpha\right|^2\right)
\Laguerre_n^{\mu-1}\!\left(-\frac{\eta}
{\frac{\bar{N}_\mathrm{nc}}{\mu}
\left(1+\frac{\bar{N}_\mathrm{nc}}{\mu}\right)}\left|\alpha\right|^2\right).\nonumber
\end{eqnarray}
In the operator form, the POVM is written as
\begin{eqnarray}
&&\hat{\Pi}_n\!\left(\bar
N_\mathrm{nc},\mu\right)=\frac{\left(\frac{\bar{N}_\mathrm{nc}}{\mu}\right)^n}
{\left(1+\frac{\bar{N}_\mathrm{nc}}{\mu}\right)^{n+\mu}}
\label{POVMMultimodeOperator}\\
&&\times\!:\exp\left(-\frac{\eta}{1+\frac{\bar{N}_\mathrm{nc}}{\mu}}\hat{a}^\dag\hat{a}\right)
\Laguerre_n^{\mu-1}\!\left(-\frac{\eta}
{\frac{\bar{N}_\mathrm{nc}}{\mu}
\left(1+\frac{\bar{N}_\mathrm{nc}}{\mu}\right)}\hat{a}^\dag\hat{a}\right):.\nonumber
\end{eqnarray}

One can prove that in the framework of the considered model,
Eqs.~(\ref{POVMMultimodeSymbol}) and (\ref{POVMMultimodeOperator})
can also be applied for the detection of wideband quantum light even
in the case in which several modes of the heat bath are coupled to
the radiation field. In this case, the operator $\hat{a}$ describes
the corresponding nonmonochromatic mode. Therefore, we conclude that
the statistics of photocounts does not depend on the number $\nu$ of
thermal-bath modes coupled to the signal and depends only on the
total number $\mu$ of heat-bath modes.

In most practical implementations of the quantum optical schemes,
one seemingly deals with a large number of thermal modes. If, for
example, the bandwidth of the heat bath includes the whole optical
band, i.e., $\Delta\omega\sim 10^{15}\, \textmd{s}^{-1}$, and the
detection time is $T=10^{-9}\,\textmd{s}$, from Eq.~(\ref{Mu}) it
follows that the number of thermal modes $\mu\sim 10^{6}$. The limit
of Eq.~(\ref{POVMMultimodeSymbol}) for a large number of noise
modes,
\begin{equation}
{\Pi}_n\!\left(\alpha;\bar
N_\mathrm{nc}\right)=\lim_{\mu\rightarrow+\infty}{\Pi}_n\!\left(\alpha;\bar
N_\mathrm{nc},\mu\right),
\end{equation}
can be easily obtained as
\begin{equation}
\Pi_n\left(\alpha;\bar N_\mathrm{nc}\right)=
\frac{\left(\eta\left|\alpha\right|^{2}+\bar{N}_\mathrm{nc}\right)^n}{n!}
\exp\left(-\eta\left|\alpha\right|^2-\bar{N}_\mathrm{nc}\right).\label{POVMLimitSymbol}
\end{equation}
This equation determines the normal-ordered symbol of the POVM in
the presence of noise that can be applied in most practical
situations. The corresponding operator form of the POVM is
represented as
\begin{equation}
\hat{\Pi}_n\left(\bar N_\mathrm{nc}\right)=
:\frac{\left(\eta\,\hat{a}^\dag\hat{a}\,+\bar{N}_\mathrm{nc}\right)^n}{n!}
\exp\left(-\eta\,\hat{a}^\dag\hat{a}-\bar{N}_\mathrm{nc}\right):.\label{POVMLimitOperator}
\end{equation}
Coupling between the noise modes and the radiation field is
negligible as soon as
\begin{equation}
\frac{\bar{N}_\mathrm{nc}}{\mu}\ll 1.\label{CondPoisson}
\end{equation}

Consider the variance of photocounts, $\overline{\Delta n^2}$,
\begin{equation}
\overline{\Delta
n^2}=\bar{n}+\eta^2\left\langle:\Delta\hat{n}^2:\right\rangle+
\frac{\bar{N}_\mathrm{nc}}{\mu}\left(2\eta\left\langle
\hat{n}\right\rangle+\bar{N}_\mathrm{nc}\right).\label{ShotExcessNoise}
\end{equation}
In this expression,
\begin{equation}
\bar{n}=\eta\left\langle\hat{n}\right\rangle+\bar{N}_\mathrm{nc}\label{FirstMomentPhotnNumber}
\end{equation}
is the mean number of photocounts, which includes the mean number of
photons, $\left\langle\hat{n}\right\rangle$, and that of noise
counts, $\bar{N}_\mathrm{nc}$. This means that noise counts
contribute to the shot noise of the detector. The second term in
Eq.~(\ref{ShotExcessNoise}) describes excess noise caused by the
stochastic nature of the light, and
$\left\langle:\Delta\hat{n}^2:\right\rangle$ is the normal-ordered
dispersion of photon number.

The third term in Eq.~(\ref{ShotExcessNoise}) corresponds to the
excess noise caused by noise counts. This noise disappears when
condition (\ref{CondPoisson}) is satisfied. Therefore, for the
detection times
\begin{equation}
T \approx \frac{\bar{N}_\mathrm{nc}}{\Delta
\omega},\label{DetectionTime}
\end{equation}
the statistics of noise counts differs from Poissonian, and the
corresponding POVM is described by
Eq.~(\ref{POVMMultimodeOperator}). This can take place, e.g., for
the femtosecond detection times. However, such an interesting case
from the point of different applications cannot be implemented with
modern technologies. In this case the contribution of
$\bar{N}_\textrm{nc}^2/\mu$ into the third term is significant in
any situation, and another contribution, $2\eta\left\langle
\hat{n}\right\rangle\bar{N}_\mathrm{nc}/\mu$, is significant only in
the case in which the field source of noise counts is coupled to the
radiation field. For detection times sufficiently greater than that
defined by Eq.~(\ref{DetectionTime}), noise counts obey the
Poissonian statistics and contribute only to the shot noise of the
detector, cf. Eq.~(\ref{FirstMomentPhotnNumber}). The corresponding
POVM is described by Eq.~(\ref{POVMLimitOperator}).

As an example, consider the light with sub-Poissonian statistics of
photocounts \cite{Short}. This property can be characterized by the
Mandel parameter \cite{MandelBook, Mandel4}, which is defined as the
ratio of the excess-noise and shot-noise variations. The Mandel
parameter in the case of Poissonian statistics of noise counts,
\begin{equation}
Q=\eta\frac{\left\langle:\Delta\hat{n}^2:\right\rangle}
{\left\langle\hat{n}\right\rangle+{\displaystyle\frac{\bar{N}_\mathrm{nc}}{\eta}}},
\end{equation}
is a monotonic function of $\bar{N}_\mathrm{nc}/\eta$. For large
values of $\bar{N}_\mathrm{nc}$, the Mandel parameter slowly tends
to zero. Otherwise, in the case of the detection time comparable
with that given by Eq.~(\ref{DetectionTime}), the Mandel parameter
as a function of $\bar{N}_\mathrm{nc}$,
\begin{equation}
Q=\eta\frac{\left\langle:\Delta\hat{n}^2:\right\rangle+{\displaystyle\frac{1}{\mu}
\frac{\bar{N}_\mathrm{nc}}{\eta}\left(2\left\langle
\hat{n}\right\rangle+\frac{\bar{N}_\mathrm{nc}}{\eta}\right)}}
{\left\langle\hat{n}\right\rangle+{\displaystyle\frac{\bar{N}_\mathrm{nc}}{\eta}}},
\end{equation}
has a threshold value,
\begin{equation}
\bar{N}_\mathrm{nc}=\eta\sqrt{\left\langle\hat{n}\right\rangle^2
-\mu\left\langle:\Delta\hat{n}^2:\right\rangle}-\eta\left\langle\hat{n}\right\rangle,
\label{Threshold}
\end{equation}
starting from which the Mandel parameter is positive and the
corresponding statistics is super-Poissonian.

In the ideal photodetection, the presence of $n$ photons is always
converted to the $n$ photocounts. In other words, for the
Fock-number state, $\ket{n}$, the probability to get $n$ photocounts
is always equal to 1. In the case of noisy detection, the presence
of $n$ photons may result in a different number of photocounts, $m$.
The probability to get $m$ photocounts under the condition that $n$
photons are present is
\begin{equation}
  P_{m|n}=\bra{n}\hat{\Pi}_m\ket{n}. \label{CondProb}
\end{equation}
The POVM is expanded into series,
\begin{equation}
\hat{\Pi}_m=\sum\limits_{n=0}^{+\infty}P_{m|n}\ket{n}\bra{n},
\end{equation}
and the probability to get $m$ photocounts, $P_m$, is
\begin{equation}
P_m=\sum\limits_{n=0}^{+\infty}P_{m|n}\,
p_n,\label{ProbabilitySeries}
\end{equation}
where
\begin{equation}
p_n=\bra{n}\hat\varrho\ket{n}
\end{equation}
is the noiseless statistics of photocounts, i.e., the probability
that $n$ photons are present.

For the case of Poissonian statistics of noise counts, the
conditional probability, Eq.~(\ref{CondProb}), is given by
\begin{equation}
P_{m|n}\left(\eta,\bar{N}_\mathrm{nc}\right)=
e^{-\bar{N}_\mathrm{nc}}
\bar{N}_\mathrm{nc}^{m-n}\eta^{n}\frac{n!}{m!}\Laguerre_n^{m-n}\!
\left(\frac{\bar{N}_\mathrm{nc}(\eta-1)}{\eta}\right)\label{CondProbPois1}
\end{equation}
for $m>n$ and
\begin{equation}
P_{m|n}\left(\eta,\bar{N}_\mathrm{nc}\right)=
e^{-\bar{N}_\mathrm{nc}}(1-\eta)^{n-m}\eta^{m}\Laguerre_m^{n-m}\!
\left(\frac{\bar{N}_\mathrm{nc}(\eta-1)}{\eta}\right)\label{CondProbPois2}
\end{equation}
for $m\leq n$. We will consider two important limiting forms of
these expressions.

The first limit corresponds to the detectors without noise counts,
$\bar{N}_\mathrm{nc}=0$,
\begin{equation}
P_{m|n}\left(\eta,\bar{N}_\mathrm{nc}=0\right)=0
\end{equation}
for $m>n$ and
\begin{equation}
P_{m|n}\left(\eta,\bar{N}_\mathrm{nc}=0\right)={n\choose m}
\eta^{m}(1-\eta)^{n-m}
\end{equation}
for $m\leq n$. Since in this case Eq.~(\ref{ProbabilitySeries})
presents the binomial transform, it can be analytically inverted,
\begin{equation}
p_n=\sum\limits_{m=n}^{+\infty}{m\choose n}
\frac{1}{\eta^n}\left(1-\frac{1}{\eta}\right)^{m-n}
P_m,\label{Inverse1}
\end{equation}
by replacing $\eta$ with $1/\eta$ \cite{Vogel}.

Another limit corresponds to the detectors with noise counts, and
with unit efficiency, $\eta=1$,
\begin{equation}
P_{m|n}\left(\eta=1,\bar{N}_\mathrm{nc}\right)=e^{-\bar{N}_\mathrm{nc}}
\frac{\bar{N}_\mathrm{nc}^{m-n}}{(m-n)!}
\end{equation}
for $m\geq n$ and
\begin{equation}
P_{m|n}\left(\eta=1,\bar{N}_\mathrm{nc}\right)=0
\end{equation}
for $m<n$. This is the shifted Poisson distribution. For this case,
Eq.~(\ref{ProbabilitySeries}) can also be analytically inverted,
\begin{equation}
p_n=e^{\bar{N}_\mathrm{nc}}\sum\limits_{m=0}^{n}
\frac{(-\bar{N}_\mathrm{nc})^{n-m}}{(n-m)!} P_m.\label{Inverse2}
\end{equation}
Equations~(\ref{Inverse1}) and (\ref{Inverse2}) express in an
explicit form the noiseless statistics in terms of the noisy
statistics of photocounts for two special cases once characteristics of noise are known.

As has been mentioned above, the POVM has the form of
Eq.~(\ref{POVMMultimodeOperator}) for the sufficiently small
detection times defined by Eq.~(\ref{DetectionTime}). In this case,
the conditional probability, Eq.~(\ref{CondProb}), is given by
\begin{eqnarray}
&&P_{m|n}\left(\eta,\bar{N}_\mathrm{nc},\mu\right)=
\frac{\left(\frac{\bar{N}_\mathrm{nc}}{\mu}\right)^{m}
\left(1+\frac{\bar{N}_\mathrm{nc}}{\mu}-\eta\right)^{n}}
{\left(1+\frac{\bar{N}_\mathrm{nc}}{\mu}\right)^{n+m+\mu}}\\
&&\times{m+\mu-1\choose m}
\sideset{_2}{_1}\HyperGeom\left(-n,-m;\mu;\frac{\eta}
{\frac{\bar{N}_\mathrm{nc}}{\mu}
\left(1+\frac{\bar{N}_\mathrm{nc}}{\mu}-\eta\right)}\right),\nonumber
\end{eqnarray}
where $\sideset{_2}{_1}\HyperGeom\left(m,n;k;z\right)$ is the
hypergeometric function. It can be checked by direct calculations
that for $\mu\gg \bar{N}_\mathrm{nc}$ this equation is transformed
into Eqs.~(\ref{CondProbPois1}) and (\ref{CondProbPois2}).

In conclusion, we note that the appearance of dark counts and
background radiation noise can be ascribed to a heat bath of
harmonic oscillators -- thermal noise modes of different nature;
each of them is either coupled or uncoupled to the signal. For the
realistic situation of a large number of noise modes, noise counts
are not coupled to the signal field and obey the Poissonian
statistics. Their contribution is totally accounted for as the shot
noise. In this case, nonclassicality of quantum-light statistics
disappears slowly as the total number of noise counts increases.

\acknowledgements

The authors gratefully acknowledge the Fundamental Researches State
Fund of Ukraine for supporting this work. A.A.S. also thanks NATO
for financial support.


\begin{thebibliography}{99}
\bibitem{Mandel1} L. Mandel, E.C.G. Sudarshan, and E. Wolf, Proc.
Phys. Soc. (London) {\bf 84}, 435 (1964).
\bibitem{MandelBook} L. Mandel and E. Wolf, {\it Optical Coherence and Quantum Optics},
(Cambridge University Press, 1995).
\bibitem{Glauber1} R.J. Glauber, Phys. Rev. {\bf 130}, 2529 (1963)
{\bf 131}, 2766 (1963).
\bibitem{Glauber2} R.J. Glauber, Phys. Rev. {\bf 131}, 2766 (1963).
\bibitem{Kelley} P.L. Kelley and W.H. Kleiner, Phys. Rev. {\bf 136},
316 (1964).
\bibitem{VogelBook} W. Vogel and D.-G. Welsch, {\it Quantum
Optics}, (Wiley--VCH, Berlin, 2006).
\bibitem{Waks} E. Waks, E. Diamanti, B.C. Sanders, S.D.
Bartlett, and Y. Yamamoto, Phys. Rev. Lett. {\bf 92}, 113602 (2004).
\bibitem{Takeuchi} S. Takeuchi, J. Kim, Y. Yamamoto, and H.H. Hogue,
Appl. Phys. Lett. {\bf 74}, 1063 (1999).
\bibitem{Karp} S. Karp, E.L.
O'Neill, and R.M. Gagliardi, Proc. IEEE {\bf 58}, 1611 (1970).
\bibitem{Pratt} W.K. Pratt, {\it Laser Communication Systems}, (Wiley
and Sons, NY, 1969).
\bibitem{Rohde} P.P. Rohde and T.C. Ralph, J. Mod. Opt. {\bf 53},
1589 (2006); M.G.A. Paris, Phys. Lett. A {\bf 289}, 167 (2001).
\bibitem{Short} R. Short and L. Mandel, Phys. Rev. Lett. {\bf
51}, 384 (1983).
\bibitem{Slusher} R.E. Slusher, L.W. Hollberg, B. Yurke, J.C. Mertz, and
J.F Valley, Phys. Rev. Lett. {\bf 55}, 2409 (1985).
\bibitem{Busch} P. Busch, P.J. Lahti, and P. Mittelstaedt, {\it The Quantum Theory
of Measurement}, (Springer, Berlin, 1996).
\bibitem{Wallentowitz} S. Wallentowitz and W. Vogel, Phys. Rev. A {\bf 53}, 4528
(1996).
\bibitem{VogelPaper} W. Vogel and J. Grabow, Phys. Rev. A {\bf 47},
4227 (1993).
\bibitem{Vogel} D.-G. Welsch, W. Vogel, T. Opartn\'y, Progr. Opt. {\bf
39}, 63 (1999).
\bibitem{HusimiKano} K. Husimi, Proñ. Phys. Math. Soc. Japan {\bf 22}, 264
(1940); Y. Kano, J. Math. Phys. {\bf 6}, 1913 (1965).
\bibitem{Mandel2} L. Mandel, Proc. Phys. Soc. (London) {\bf 74}, 233 (1959).
\bibitem{Mandel3} L. Mandel, Proc. Phys. Soc. (London) {\bf 81}, 1104 (1963).
\bibitem{Saleh} B. Saleh, {\it Photoelectron Statistics},
(Springer-Verlag, Berlin, 1978).
\bibitem{Mandel4} L. Mandel, Opt. Lett. {\bf 4}, 205 (1979).
\end{thebibliography}
\end{document}